\documentclass[aps,prb,twocolumn,amssymb,floatfix]{revtex4}  

\usepackage{graphicx}

\renewcommand{\vec}[1]{{\mathbf #1}}

\begin{document}

\title{Instability of speckle patterns in random media\\
with noninstantaneous Kerr nonlinearity}

\author{S.E. Skipetrov}

\affiliation{Laboratoire de Physique et Mod\'elisation des Milieux Condens\'es\\
CNRS, 38042 Grenoble, France}

\begin{abstract}
Onset
of the instability of a multiple-scattering speckle pattern
in a random medium with Kerr nonlinearity is significantly affected
by the noninstantaneous character of the nonlinear medium response. The fundamental
time scale of the spontaneous speckle dynamics beyond the instability
threshold is set by the largest of times $T_{\mathrm{D}}$ and
$\tau_{\mathrm{NL}}$, where $T_{\mathrm{D}}$ is the time required for the
multiple-scattered waves to propagate through the random sample and
$\tau_{\mathrm{NL}}$
is the relaxation time of the nonlinearity. Inertial nature of the nonlinearity
should complicate the experimental observation of the instability
phenomenon.\\

\begin{center}
Accepted for publication in \textit{Optics Letters} \copyright~2003 Optical Society of America, Inc.
\end{center}
\end{abstract}

\maketitle 

Instabilities, self-oscillations and chaos are widely encountered in nonlinear
optical systems \cite{gibbs85,agrawal95,arrechi99}.
In particular, the so-called modulation instability (MI), consisting in the growth of small perturbations of the
wave amplitude and phase, resulting from the combined effect of nonlinearity and diffraction,
can be observed for both coherent \cite{agrawal95,shih02} and incoherent \cite{soljacic00,kip00,klinger01,chen02,hall02} light.
In the latter case, propagation of a powerful,
spatially incoherent quasimonochromatic light beam through an optically
\textit{homogeneous} nonlinear crystal results in ordered
or disordered spatial patterns \cite{soljacic00,kip00,klinger01,chen02,hall02}.
Recently, instabilities have also been predicted \cite{skip00,skip03} for diffuse light in
\textit{random} media with Kerr nonlinearity (sample size $L \gg$ mean free path $\ell \gg$
wavelength $\lambda$). The strength
of the nonlinearity has to exceed a threshold for the instability to develop, similarly to the case of incoherent MI
(coherent MI has no threshold), but in contrast to the latter,
(a) initially coherent wave loses its spatial coherence
\textit{inside} the medium, due to the multiple scattering on heterogeneities of the refractive index,
(b) temporal coherence of the
incident light is assumed to be perfect (in the case of incoherent MI,
the coherence time of the incident beam is much shorter than the response time of the medium),
(c) incident light beam is completely destroyed by the scattering after a distance 
$\sim \ell$  and light propagation is diffusive in the bulk of the sample, and
(d) scattering provides a distributed feedback mechanism, absent in the case of MI.

Since the noninstantaneous nature of nonlinearity proved to be important for the development of optical
instabilities \cite{gibbs85,agrawal95,arrechi99,shih02,soljacic00,kip00,klinger01,chen02,hall02},
it is natural to
consider its effect on the instability of diffuse waves in random nonlinear media.
Up to now, the latter phenomenon has only been studied assuming the instantaneous medium response \cite{skip00,skip03}.
In the present letter, we show that while the absolute
instability threshold does not depend on the nonlinearity response time $\tau_{\mathrm{NL}}$,
the dynamics of the speckle pattern above the threshold is
extremely sensitive to the relation between $\tau_{\mathrm{NL}}$ and the time
$T_{\mathrm{D}} = L^2/D$ that takes the diffuse wave to propagate through the random
sample ($D = c \ell / 3$ is the diffusion constant and $c$ is the speed
of light in the average medium). We derive explicit expressions for
the characteristic time scale $\tau$ of spontaneous intensity fluctuations beyond the instability
threshold and the maximal
Lyapunov exponent $\Lambda_{\mathrm{max}}$, characterizing the predictability of the
system behavior. Finally, we discuss the experimental implications of our results.

\begin{figure}
\includegraphics[width=5cm]{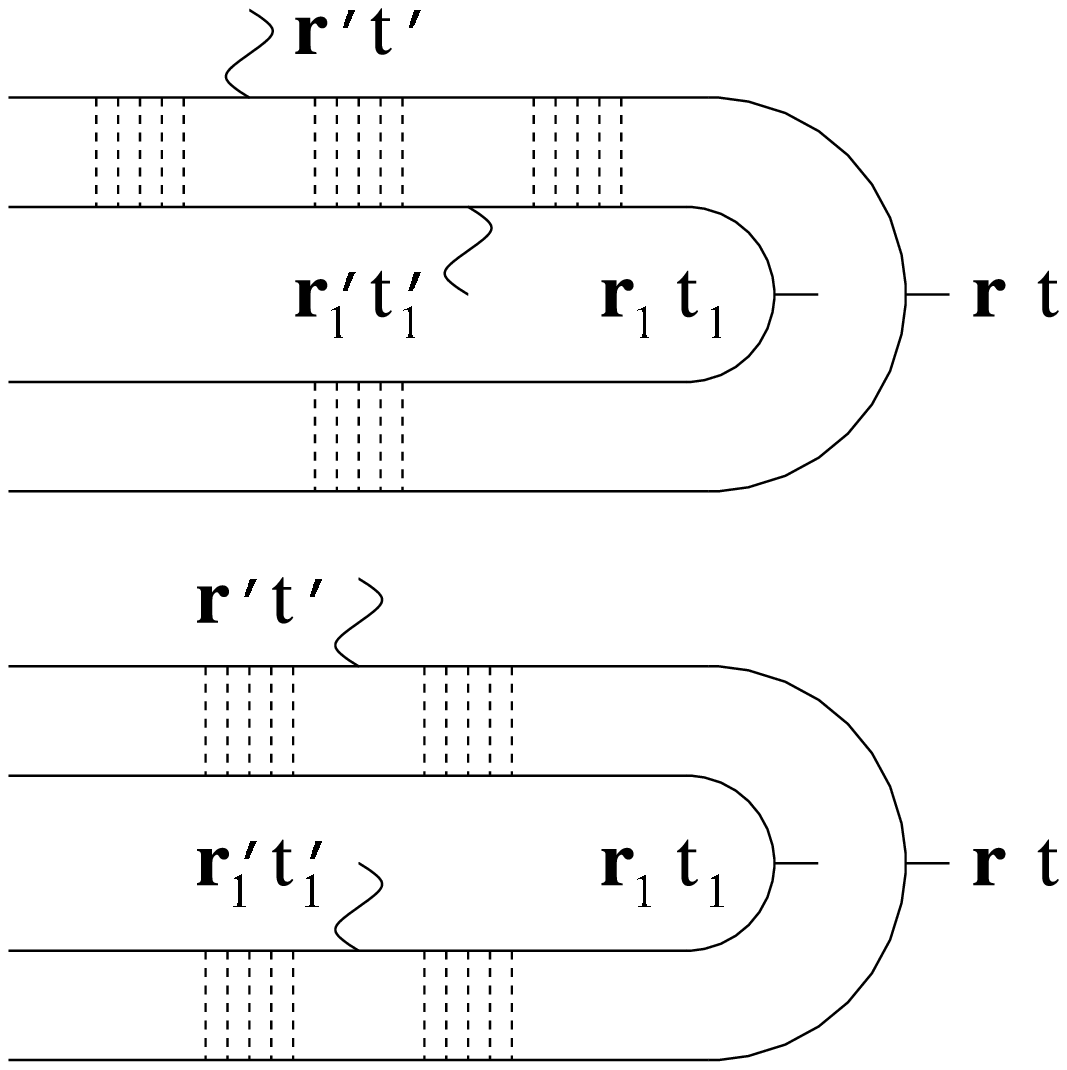}
\vspace*{5mm}
\caption{Diagrams contributing to the correlation function
$\left< q^{(i)}( \vec{r}, \vec{r}^{\, \prime}, t-t^{\, \prime})
q^{(j)*}( \vec{r}_1, \vec{r}_1^{\, \prime}, t_1-t_1^{\, \prime}) \right>$
of the random response functions $\vec{q}$ entering into
Eq.\ (\ref{djnl}). Solid lines represent the retarded and advanced Green's functions
of the linear wave equation, dashed lines denote scattering of the two
connected wave fields on the same heterogeneity. Wavy lines are $k_0^2$ vertices.}
\label{fig1}
\end{figure}

We consider a plane monochromatic wave of frequency $\omega_0$ incident on a
random sample of typical size $L \gg \ell$. The real dielectric constant of the sample
can be written as $\varepsilon(\vec{r}, t) =
\varepsilon_0 + \delta \varepsilon(\vec{r}) + \Delta \varepsilon_{\mathrm{NL}}(\vec{r}, t)$,
where $\varepsilon_0$ and $\delta \varepsilon(\vec{r})$ 
are the average value and the fluctuating part of the linear
dielectric constant, respectively, and $\Delta \varepsilon_{\mathrm{NL}}(\vec{r}, t)$ denotes
the nonlinear part of $\varepsilon$.
Without loss of generality, we assume $\varepsilon_0 = 1$ and 
$\left< \delta \varepsilon(\vec{r}) \delta \varepsilon(\vec{r}_1) \right> =
4 \pi/(k_0^4 \ell) \delta(\vec{r}-\vec{r}_1)$.
Under conditions of weak scattering $k_0 \ell \gg 1$ (where
$k_0 = \omega_0/c$),
the average intensity $\left< I(\vec{r}) \right>$
of the scattered wave inside the sample can be described by a diffusion equation,
the short-range correlation function of intensity fluctuations
$\delta I(\vec{r}, t) = I(\vec{r}, t) - \left< I(\vec{r}) \right>$ decreases to zero at distances
of order $\lambda$,
and the long-range correlation of $\delta I(\vec{r}, t)$
can be found from the Langevin equation \cite{zyuzin87} (see Ref.\ \onlinecite{rossum99} for a review
of wave diffusion in random media):
\begin{eqnarray}
\frac{\partial}{\partial t} \delta I(\vec{r}, t)  - D \nabla^2  \delta I(\vec{r}, t) =
- \vec{\nabla} \cdot {\vec j}_\mathrm{ext}(\vec{r}, t).
\label{langevin}
\end{eqnarray}
In a nonlinear medium, the random external Langevin currents
${\vec j}_\mathrm{ext}(\vec{r}, t)$ obey a dynamic equation \cite{skip03}
\begin{eqnarray}
\frac{\partial}{\partial t} \vec{j}_\mathrm{ext} (\vec{r}, t) &=& 
\int_V \mathrm{d}^3 \vec{r}^{\, \prime} \int_{0}^{\infty} \mathrm{d} \Delta t \,
\vec{q}( \vec{r}, \vec{r}^{\, \prime}, \Delta t)
\nonumber \\
&\times& \frac{\partial}{\partial t} \Delta \varepsilon_{\mathrm{NL}}(\vec{r}^{\, \prime}, t - \Delta t),
\label{djnl}
\end{eqnarray}
where the integration is over the volume $V$ of the random medium
and $\vec{q}( \vec{r}, \vec{r}^{\, \prime}, \Delta t)$
is a random, sample-specific response function with zero average and a correlation
function given by the diagrams of Fig.\ \ref{fig1}.

Supplemented by a Debye-relaxation
relation for $\Delta \varepsilon_{\mathrm{NL}}(\vec{r}, t)$:
\begin{eqnarray}
\tau_{\mathrm{NL}} \frac{\partial}{\partial t} \Delta \varepsilon_{\mathrm{NL}}(\vec{r}, t) =
-\Delta \varepsilon_{\mathrm{NL}}(\vec{r}, t) + 2 n_2 I(\vec{r}, t),
\label{debye}
\end{eqnarray}
where $n_2$ is the nonlinear coefficient,
Eqs.\ (\ref{langevin}) and (\ref{djnl}) can now be used to perform
a linear stability analysis of the multiple-scattering speckle pattern in a
random medium with noninstantaneous nonlinearity.
Assuming $\delta I(\vec{r}, t) = \delta I(\vec{r}, \alpha) \exp(\alpha t)$ with
$\alpha = i \Omega + \Lambda$ ($\Omega > 0$),
and similarly for $\vec{j}_\mathrm{ext} (\vec{r}, t)$
and $\Delta \varepsilon_{\mathrm{NL}}(\vec{r}, t)$,
we use Eqs.\ (\ref{djnl}) and (\ref{debye}) to express the correlation function
of Langevin currents
$\left< j_\mathrm{ext}^{(i)} (\vec{r}, \alpha) j_\mathrm{ext}^{(j)*} (\vec{r}_1, \alpha)
\right>$ in terms of the intensity correlation function
$\left< \delta I (\vec{r}, \alpha) \delta I^* (\vec{r}_1, \alpha) \right>$
and Eq.\ (\ref{langevin}) to express the latter correlation function in terms
of the former.
For large enough sample size $L/\ell \gg k_0 \ell$ and moderate
frequency $\Omega T_{\mathrm{D}} \ll [L/(k_0 \ell^2)]^2$ we can neglect the short-range
correlation of intensity fluctuations and write
the condition
of consistency of the two obtained equations [and hence the condition of
consistency of Eqs.\ (\ref{langevin}--\ref{debye})] as
\begin{eqnarray}
p \simeq F(\Omega T_{\mathrm{D}}, \Lambda T_{\mathrm{D}})
H(\Omega \tau_{\mathrm{NL}}, \Lambda \tau_{\mathrm{NL}}),
\label{pp}
\end{eqnarray}
where a numerical factor of order unity has been omitted,
$p = \left< \Delta n_{\mathrm{NL}} \right>^2 (L/\ell)^3$
is the effective nonlinearity parameter,
$\left< \Delta n_{\mathrm{NL}} \right> = n_2 \left< I(\vec{r}) \right>$ is the average value of the
nonlinear part of refractive index,
and $\left< I(\vec{r}) \right>$ is assumed to be approximately constant
inside the sample. The function $F$ in Eq.\ (\ref{pp})
is the same as in the case of instantaneous nonlinearity \cite{skip03},
while $H(x, y) = x^2 + (1+y)^2$ originates from the noninstantaneous nature of the
nonlinear response. In the limit of $\tau_{\mathrm{NL}} \rightarrow 0$ (instantaneous
nonlinearity), $H \equiv 1$ and Eq.\ (\ref{pp}) reduces to Eq.\ (7) of Ref.\ \onlinecite{skip03}.

\begin{figure}
\begin{center}
\hspace*{-1.5cm}\includegraphics[width=8cm]{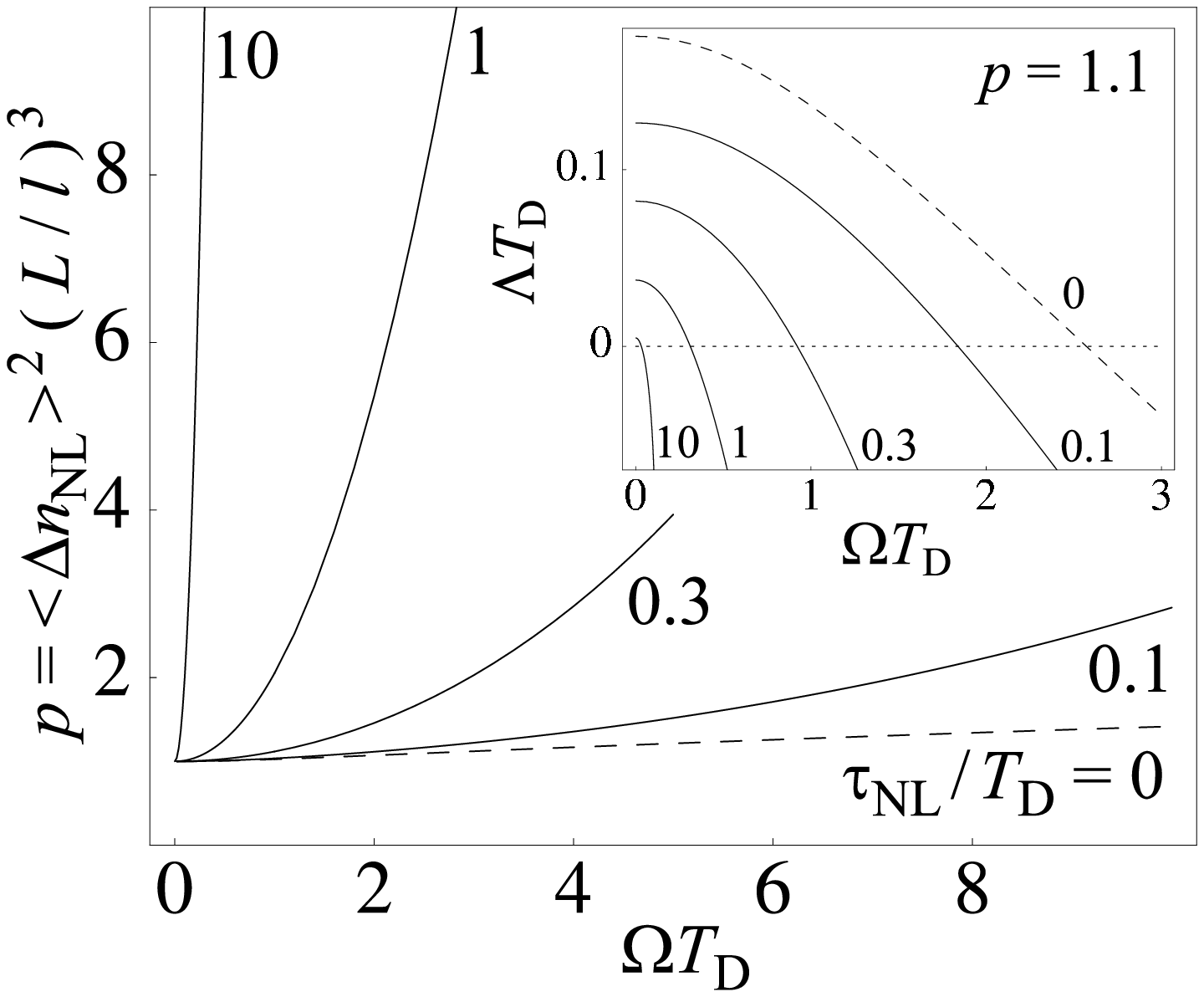}
\vspace*{-8cm}
\caption{Instability threshold as a function of the
excitation frequency $\Omega$ (in units of $T_{\mathrm{D}}^{-1}$) for several ratios of the
nonlinearity relaxation time $\tau_{\mathrm{NL}}$ and the time
$T_{\mathrm{D}}$ required for light to diffuse through the random sample.
Dashed line corresponds to
$\tau_{\mathrm{NL}}/T_{\mathrm{D}} = 0$.
Inset: Lyapunov exponent $\Lambda$ vs. $\Omega$ (both in units of $T_{\mathrm{D}}^{-1}$) slightly above
the threshold ($p = 1.1$) for the same values of $\tau_{\mathrm{NL}}/T_{\mathrm{D}}$ as
in the main plot.}
\label{fig2}
\end{center}
\end{figure}

For given $\Omega$ and $p$, Eq.\ (\ref{pp}) determines the corresponding
value of the Lyapunov exponent $\Lambda$.
$\Lambda > 0$ signifies the instability of the speckle pattern
with respect to the excitations at frequency $\Omega$. If $p < 1$,
then $\Lambda < 0$ for all $\Omega$. As $p$ exceeds $1$, $\Lambda$ becomes positive
inside some frequency band $(0, \Omega_{\mathrm{max}})$ (see the inset of Fig.\ \ref{fig2}).
We show the frequency-dependent threshold following from Eq.\ (\ref{pp})
in Fig.\ \ref{fig2} for different values of
$\tau_{\mathrm{NL}}/T_{\mathrm{D}}$.
The threshold value of $p$ at $\Omega \gtrsim \tau_{\mathrm{NL}}^{-1}$ deviates from the result obtained
for the instantaneous nonlinearity (dashed line in Fig.\ \ref{fig2}),
while the \textit{absolute} threshold is $p = 1$ at
$\Omega = 0$, independent of $\tau_{\mathrm{NL}}/T_{\mathrm{D}}$.

To study the onset of the instability, we consider the limit of $0 < p - 1 \ll 1$
in more detail. In this limit, a series expansion of the function $F$ in
Eq.\ (\ref{pp}) can be found: $F(x, y) \simeq 1 + C_1 x^2
+ y (C_2 - C_3 x^2)$, where we keep only the terms quadratic in
$x$ and linear in $y$, and the numerical constants are
$C_1 \simeq 0.031$, $C_2 \simeq 0.59$, $C_3 \simeq 0.045$. At a given frequency
$\Omega$,
the threshold value of $p$ then becomes
$p \simeq 1 + \left[ C_1 + (\tau_{\mathrm{NL}}/T_{\mathrm{D}})^2 \right]
(\Omega T_{\mathrm{D}})^2$,
while at a given $p$ the maximum excited frequency is
$\Omega_{\mathrm{max}} = T_{\mathrm{D}}^{-1} \left[ C_1 +
(\tau_{\mathrm{NL}}/T_{\mathrm{D}})^2 \right]^{-1/2} (p-1)^{1/2}$ and
the shortest typical time of spontaneous intensity fluctuations is
$\tau = 1/\Omega_{\mathrm{max}}$.

As follows from the above analysis, a continuous low-frequency spectrum
of frequencies $(0, \Omega_{\mathrm{max}})$
is excited at $p > 1$, and the Lyapunov exponent $\Lambda$ decreases
monotonically with $\Omega$ (see the inset of Fig.\ \ref{fig2}).
This allows us to hypothesize that at $p = 1$ the speckle pattern undergoes
a transition from a stationary to chaotic state.
Such a behavior should be contrasted from the ``route to chaos'' through a sequence
of bifurcations, characteristic of many nonlinear systems \cite{gibbs85,agrawal95,arrechi99}.
Given initial conditions at $t = 0$ and deterministic dynamic equations,
the evolution of a chaotic system can only be predicted for
$t < 1/\Lambda_{\mathrm{max}}$. 
We find the maximal Lyapunov exponent 
$\Lambda_{\mathrm{max}} = T_{\mathrm{D}}^{-1}
(p-1)/(C_2 + 2 \tau_{\mathrm{NL}}/T_{\mathrm{D}})$.
It is now easy to see that the fundamental time scale of the speckle pattern
dynamics at $p > 1$ is set by the largest of the times $T_{\mathrm{D}}$ and
$\tau_{\mathrm{NL}}$. Indeed, in the case of fast nonlinearity
($\tau_{\mathrm{NL}} \ll T_{\mathrm{D}}$) we have
$\Omega_{\mathrm{max}} \propto T_{\mathrm{D}}^{-1} (p-1)^{1/2}$,
$\tau \propto T_{\mathrm{D}} (p-1)^{-1/2}$, and
$\Lambda_{\mathrm{max}} \propto T_{\mathrm{D}}^{-1}
(p-1)$.
In the opposite limit of slow nonlinear response
($\tau_{\mathrm{NL}} \gg T_{\mathrm{D}}$) we find
$\Omega_{\mathrm{max}} \propto \tau_{\mathrm{NL}}^{-1} (p-1)^{1/2}$,
$\tau \propto \tau_{\mathrm{NL}} (p-1)^{-1/2}$, and
$\Lambda_{\mathrm{max}} \propto \tau_{\mathrm{NL}}^{-1}
(p-1)$, respectively.

The origin of the instability of the speckle pattern $I(\vec{r}, t)$ in a nonlinear random medium
and its sensitivity to $\tau_{\mathrm{NL}}$
can be qualitatively understood by considering $I$
as a result of interference of many partial waves traveling along various
diffusion paths. An infinitely small perturbation of $I$ at some point $\vec{r}_0$ at
time $t_0$ diffuses throughout the medium [according to Eq.\ (\ref{langevin})] and,
after some time $\Delta t$, modifies the speckle pattern inside a sphere of
radius $R \sim (D \Delta t)^{1/2}$ around $\vec{r}_0$. Due to the nonlinearity of the medium, this
leads to a change of the nonlinear part of the dielectric constant
$\Delta \varepsilon_{\mathrm{NL}}$ inside the same sphere after a time $\sim \tau_{\mathrm{NL}}$
[according to Eq.\ (\ref{debye})] and, consequently, to a modification
of the phases of the partial diffuse waves that interfere to produce a new
speckle pattern $I(\vec{r}, t)$. If the strength of
the nonlinearity exceeds the threshold given by Eq.\ (\ref{pp}),
$I(\vec{r}_0, t) - I(\vec{r}_0, t_0)$ can be larger than the initial perturbation.
The latter is hence amplified and the speckle pattern develops spontaneous dynamics.
Noninstantaneous nature of the nonlinearity leads to the damping of the
high-frequency components in the spectrum of the initial intensity
perturbation, raising the instability threshold for $\Omega \gtrsim \tau_{\mathrm{NL}}^{-1}$, as we show
in Fig.\ \ref{fig2}. This phenomenon is similar in its origin to the increase of the speed of coherent
MI patterns with decreasing $\tau_{\mathrm{NL}}$ in a \textit{homogeneous} nonlinear medium \cite{shih02}.
Similarly to the case of coherent MI \cite{shih02}, the instability is arrested in the limit
of $\tau_{\mathrm{NL}} \rightarrow \infty$, in contrast to the case of incoherent MI, where
$\tau_{\mathrm{NL}} \rightarrow \infty$ (more precisely, $\tau_{\mathrm{NL}} \gg$ than the coherence time of the
incident beam) is an explicit assumption of the theoretical models
\cite{soljacic00,kip00,hall02}. 

To conclude, we discuss the experimental implications of our results.
$\Delta n_{\mathrm{NL}}$ up to $10^{-3}$ can be
realized in inorganic photorefractive crystals \cite{kip00,klinger01,chen02}.
If scattering centers (optical defects) are introduced in the crystal, $L/\ell \sim 10^2$
will suffice to reach the instability threshold $p \simeq 1$.
Alternatively, in nematic liquid crystals both nonlinearity ($\Delta n_{\mathrm{NL}}$ up to 0.1, see, e.g.,
Ref.\ \onlinecite{muenster97}) and scattering
($L/\ell > 10$ has been reported in Ref.\ \onlinecite{kao96}) can be sufficiently
strong to obtain $p \simeq 1$.
The nonlinearity is rather slow in the above cases,
$\tau_{\mathrm{NL}}$ exceeds $T_{\mathrm{D}}$ ($T_{\mathrm{D}} \sim 10 \div 100$ ns for
$\ell \sim 0.1 \div 1$ mm and $L \sim 1$ cm), and the results obtained in the present letter are,
therefore, of particular relevance.
Finally, although the noninstantaneous nature of
nonlinearity does not affect the absolute instability threshold, it should complicate the experimental
observation of the instability phenomenon. Indeed, in a real experiment
only sufficiently fast fluctuations can be detected.
Inasmuch as $\Omega_{\mathrm{max}}$ decreases with $\tau_{\mathrm{NL}}$, larger
$p$ (and thus stronger nonlinearities) will be required for the instability to be
detectable in a medium with noninstantaneous nonlinearity.

\end{document}